\documentclass{pasj01}

\usepackage{lscape}
\usepackage{natbib}

\begin{document}

\title{Measurement of Interstellar Polarization 
and Estimation of Galactic Extinction 
for the Direction of X-ray Black Hole Binary V404~Cygni}

\author{Ryosuke \textsc{Itoh}\altaffilmark{1,1a}, 
Yasuyuki~T. \textsc{Tanaka}\altaffilmark{2}, 
Koji~S. \textsc{Kawabata}\altaffilmark{2}, 
Makoto \textsc{Uemura}\altaffilmark{2}, 
Makoto \textsc{Watanabe}\altaffilmark{3,3a}, 
Yasushi \textsc{Fukazawa}\altaffilmark{1}, 
Yuka \textsc{Kanda}\altaffilmark{1},
Hiroshi \textsc{Akitaya}\altaffilmark{2}, 
Yuki \textsc{Moritani}\altaffilmark{2,4},
Tatsuya \textsc{Nakaoka}\altaffilmark{1},
Miho \textsc{Kawabata}\altaffilmark{1},
Kensei \textsc{Shiki}\altaffilmark{1},
Michitoshi \textsc{Yoshida}\altaffilmark{2},
Yumiko \textsc{Oasa}\altaffilmark{5} 
and Jun \textsc{Takahashi}\altaffilmark{6} 
}
\altaffiltext{1}{Department of Physical Science, Hiroshima University, Higashi-Hiroshima, Hiroshima 739-8526, Japan}
\altaffiltext{1a}{School of Science, Tokyo Institute of Technology, 2-12-1 Ohokayama, Meguro, Tokyo 152-8551, Japan; : itoh@hp.phys.titech.ac.jp}
\altaffiltext{2}{Hiroshima Astrophysical Science Center, Hiroshima University, Higashi-Hiroshima, Hiroshima 739-8526, Japan}
\altaffiltext{3}{Department of Cosmosciences, Graduate School of Science, Hokkaido University,
  Kita 8, Nishi 10, Kita-ku, Sapporo, Hokkaido 060-0810, Japan}
\altaffiltext{3a}{Department of Applied Physics, Faculty of Science, Okayama
  University of Science, 1-1 Ridai-cho, Okayama, Okayama 700-0005, Japan}
\altaffiltext{4}{5Kavli Institute for the Physics and Mathematics of the Universe (WPI), The University of Tokyo Institutes for Advanced Study, The University of Tokyo, Kashiwa, Chiba 277-8583, Japan}
\altaffiltext{5}{Faculty of Education, Saitama University, 255 Shimo-Okubo, Sakura, Saitama, Saitama 338-8570, Japan}
\altaffiltext{6}{Nishi-Harima Astronomical Observatory, Center for Astronomy, University of Hyogo, 407-2, Nishigaichi, Sayo-cho, Sayo, Hyogo 679-5313, Japan}
\maketitle

\begin{abstract} 

V404~Cygni is a well-known black hole binary candidate thought to have relativistic jets. 
It showed extreme outbursts on June 2015, characterized by a large amplitude and short 
time variation of flux in the radio, optical, and X-ray bands. 
Not only disk emission, but also synchrotron radiation from the relativistic jets were 
suggested by radio observations. 
However, it is difficult to measure the accurate spectral shape 
in the optical/near infrared band because there are uncertainties of interstellar extinction.

To estimate the extinction value for V404~Cygni, we performed photopolarimetric and 
spectroscopic observations of V404~Cygni and nearby field stars. 
Here, we estimate the Galactic extinction using interstellar polarization 
based on the observation that the origin of the optical polarization is the
interstellar medium, and 
investigate the properties of interstellar polarization around V404~Cygni. 
We found a good correlation between the color excess and polarization degree 
in the field stars. 
We also confirmed that the wavelength dependence of the polarization degree in the highly 
polarized field stars was similar to that of V404~Cygni. 
Using the highly polarized field stars, we estimated the color excess 
and the extinction, $E(B-V)=1.2 \pm 0.2$ and $3.0 < A(V) < 3.6$, respectively. 
A tendency for a bluer peak of polarization ($\lambda_{\rm max}<5500$ \AA) 
was commonly seen in the highly polarized field stars, 
suggesting that the dust grains toward this region are generally smaller than the 
Galactic average.
The corrected spectral energy distribution of V404~Cygni in the near infrared
(NIR) and optical bands 
in our results indicated a spectral break between 
$2.5 \times 10^{14}$ Hz and $3.7 \times 10^{14}$ Hz, which might be originated in
the synchrotron self absorption. 
\end{abstract}

\section{Introduction} 

The radiation from microquasars often suffers from 
Galactic extinction and polarization by the interstellar medium, 
making it difficult to determine the nature of these objects. 
Polarization in the near infrared (NIR) is one sign of synchrotron 
radiation in several microquasars \citep{2008ApJ...672..510S} and 
is an important measurement used to investigate the environment of the 
jet and the origin of variability in the microquasars. However, 
the radiation is often contaminated by interstellar polarization (ISP). 
In addition, the Galactic extinction makes it difficult to measure 
the intrinsic spectral shape of microquasars in the optical and NIR bands. 
It is important to investigate the actual properties of the interstellar medium 
to determine the precise emission model for microquasars. 

V404~Cygni (a.k.a, GS~2023+338) is a well-known X-ray black hole binary 
with $12 \pm 3 M_{\odot}$ masses \citep{1994MNRAS.271L..10S} 
at a distance of 2.4 kpc \citep{2009ApJ...706L.230M}. 
It showed extreme outbursts in June 2015, with large amplitude variability
and a time scale of seconds to hours in the radio, optical, and X-ray bands
\citep[e.g.,][]{2016Natur.529...54K}. 
In contrast to the large variation in total flux, there were very small variability 
of polarization in the optical or NIR bands.
\cite{2016ApJ...823...35T} reported no significant variability of polarization in the optical and NIR bands, while
\cite{2016MNRAS.463.1822S} and \cite{2016arXiv160802764L} detected very small polarization variation in limited
time intervals which were not covered by \cite{2016ApJ...823...35T}.
This implies that disk or optically-thick synchrotron emission (or both) dominated 
in the NIR regime. 

To obtain an accurate spectrum in the optical/NIR band, 
careful estimation of the Galactic extinction is needed. 
\cite{1993MNRAS.265..834C} performed high-resolution spectroscopic observation of V404~Cygni 
in 1991, and they estimated the spectral type of comparison star as K0(-1) III-IV. 
From this result, they determined the Galactic absorption of $A(V) = 4.0$. 
\cite{1994MNRAS.271L..10S}, however, suggested the Galactic absorption
of $A(V) < 3.3$ by NIR-band 
photometry of V404~Cygni and a tripped giant model. 

Here, we estimated the color excess of V404~Cygni 
by measuring interstellar polarization and color excess for the surrounding field stars. 
The distinctive feature of our method is that we do not require any assumption of intrinsic 
luminosity and color of comparison stars in microquasars. 
In this paragraph, we introduce the basic principles of our method. 
The relation between color excess originating in the Galactic extinction 
and ISP has been well studied \citep[e.g.,][]{2002ApJ...564..762F}. 
In general, the polarization degree (PD) of ISP increases with color excess. 
It is also known that the wavelength dependence of ISP in ultraviolet to NIR 
bands is well described by the Serkowski-law 
\citep[][details also described in Section 3.1]{1975ApJ...196..261S}. 
A wavelength at peak PD is related to the ratio of 
extinction and color excess, defined as $R_{\rm V} = A(V)/E(B-V)$, where 
$A(V)$ is extinction in the V-band and $E(B-V)$ is the color excess for $B-V$ color. 
To estimate the contribution of interstellar extinction and polarization 
for V404~Cygni, we performed optical and NIR polarimetric observations of V404~Cygni 
and nearby stars. 
First, we measured the color excess and ISP for the field stars around V404~Cygni 
and clarified the relations between these parameters. 
Then, we estimated the color excess of V404~Cygni from its PD value. 
From a measurement of the wavelength dependence of V404~Cygni, 
we estimated the $R_{\rm V}$ value with the Serkowski-law. 
Using these two parameters, $E(B-V)$ and $R_{\rm V}$, we estimated the Galactic extinction
for V404~Cygni.

\section{Observation}

\subsection{Photopolarimetric observation}

We performed {\it V-, R$_{C}$-, I$_{C}$-, J-, H-} and {\it K$_s$}-band 
photopolarimetric observation of V404~Cygni and 
field stars on 22 June 2015 using the Hiroshima Optical and Near Infrared camera 
\citep[HONIR,][]{2014SPIE.9147E..4OA} 
installed on the 1.5 m Kanata telescope, located at Higashi-Hiroshima Observatory in Japan. 
We also performed long-term photpolarimetric monitoring of V404~Cygni in the {\it R$_{\rm C}$}-band 
from 17 June 2015 to 27 September 2015 with a multi-spectral imager
\citep[MSI,][]{2012SPIE.8446E..2OW} 
installed on the 1.6 m Pirka telescope at Nayoro Observatory of Hokkaido
University in Japan. 
Each observation consisted of a sequence of exposures at four position angles 
of the achromatic half-wave plates, 0.0, 22.5, 45.0, and 67.5 degree. 
The field of view of HONIR for polarimetric observation consisted of five rectangles 
(0.8 arcmin $\times$ 10 arcmin) separated on the side by 0.8 arcmin. 
We confirmed that instrumental polarization of HONIR was less than 0.1\% in the 
optical band and 0.2\% in the NIR-band by observation of unpolarized 
standard stars \citep[HD212311 and G191B2B,][]{1992AJ....104.1563S}. 
We also corrected the instrumental depolarization by measurement of 
an artificially 100\% polarized star, with a wire-grid polarizer for each band. 
The origins of the polarization angle were calibrated with the strong polarized stars, 
HD150193 and HD19820 \citep{1992ApJ...386..562W}. 
Each measured PD and PA after all calibrations were consistent with catalog values 
of strong polarized stars, with uncertainties of $\Delta PD\sim 0.1\%$ and
$\Delta PA\sim 2$ degrees, respectively.

We used archival data of {\it Swift} observation for the 
field stars around V404~Cygni on MJD54947. 
We used the {\it B} and {\it V}-band photometric data taken by 
the UV and Optical Telescope (UVOT). 
UVOT data were reduced following the standard procedure 
for CCD photometry. 
Counts were extracted from an aperture of 5 arcsec radius 
for all filters and annulus background regions were sampled 27 arcsec away from object stars, 
and then converted to a flux using the standard zero points \citep{2008MNRAS.383..627P}.

\subsection{Spectroscopic observation}
To determine the spectral type of the field stars, 
we performed low-resolution spectroscopic observations of field stars with 
Hiroshima One-shot Wide-field Polarimeter 
\citep[HOWPol, 0.41-0.94 $\mu$m, $R=\lambda/\Delta\lambda\sim400$,][]{2008SPIE.7014E.151K} 
on the Kanata Telescope from 24 to 30 July 2015.
To investigate the relation between PD and color excess, we selected four bright
field stars, with weak to strong polarization ($0.1\% < PD < 6\%$ in the $R_{C}$-band),
for optical spectroscopic observation.
In addition, we intensively performed spectroscopic observation of all
field stars which have strong polarization ($PD>7\%$ in the $R_{C}$-band)
in order to securely compare with V404~Cyg.
In total we obtained seven spectroscopic observation of field stars
with weak to strong polarization ($0.1\% < PD < 8\%$ in the $R_{C}$-band).
Table \ref{tab:FS} shows the positions of field stars. 
The typical total exposure time for each star was about 30 minutes. 
The calibration of wavelength was performed with atmospheric emission lines for each frame. 
The flux was calibrated using observations of a spectrophotometric 
standard star (HR7596) obtained on the same nights.

\begin{table}[!htb]
  \caption{List of Field star around V404~Cygni}
  \centering
  \label{tab:FS}
  \begin{tabular}{lc}\hline\hline
    Name$^1$ & Coordinate$^2$ \\ \hline
    FS1  & 20:24:19.5 +33:52:42.3 \\
    FS2  & 20:24:11.9 +33:49:11.5 \\
    FS3  & 20:23:56.0 +33:53:28.7 \\
    FS4  & 20:23:57.1 +33:52:39.0 \\
    FS5  & 20:23:56.4 +33:48:16.9 \\
    FS6  & 20:23:49.2 +33:50:08.2 \\
    FS7  & 20:23:49.7 +33:48:23.5 \\
    V404~Cygni & 20:24:03.8 +33:52:02.2 \\
    \hline
  \end{tabular}
  \\
    {\footnotesize 1: Name in this paper. Their positions are shown in Fig. \ref{fig:Polmap},
      2:Right ascension and declination (J2000)}
\end{table}

\section{Results}

\subsection{Interstellar polarization}

\begin{figure}[!htb]
  \centering
  \includegraphics[angle=0,width=8cm]{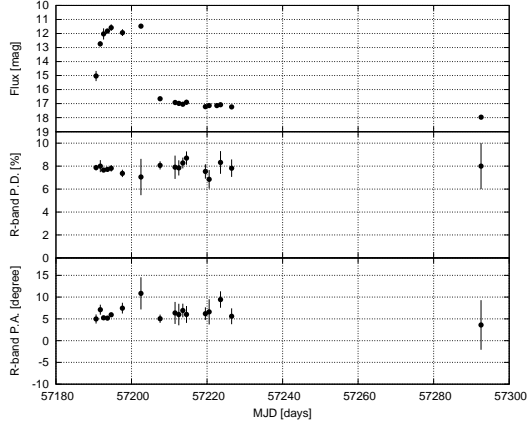}
  \caption{Long-term  light curve and temporal variability of $R_{C}$-band polarization
    of V404~Cygni. Top panel shows $R_{\rm C}$-band light curve.
    2nd panel shows temporal variability of polarization
    degree and bottom panel shows
    temporal variability of polarization angle.}
  \label{fig:LCPD}
\end{figure}

\cite{2016ApJ...823...35T} reported no temporal variability of optical linear 
polarization for V404~Cygni, although it showed extreme flares. 
Figure \ref{fig:LCPD} shows $R_{\rm C}$-band light curve and temporal measurement of polarization
for V404~Cygni from June to September 2015.
In order to investigate the variability of PD, we adopted the constant fitting for the
temporal variability of PD.
Averatge value of PD is $7.8 \pm 0.1\%$ with $\chi^2_{\nu}$ value of $\chi^2_{\nu}/{\rm d.o.f} = 9.40/16$
which corresponds to the p-value of 0.89.
It implied that there are no significant variability of PD between the active and quiescent states.

\begin{figure}[!htb]
  \centering
  \includegraphics[angle=0,width=8cm]{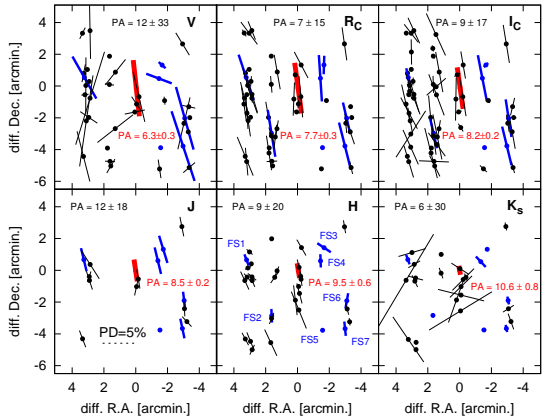}
  \caption{Interstellar polarization map around V404 Cygni in the optical in the near
    infrared band. 
    Length of bar shows polarization degree (scales are also shown in the left bottom panel)
    and direction of bar shows polarization angle on the celestial sphere.
    Red data and text indicates the data of polarization for V404~Cygni.
    Blue data indicate the data of of field stars (FS1 to 7, also shown in center bottom panel).
    Black data point indicates the data of polarization for the other field stars in the field of
    view.
    Left top text shows average value of polarization angle of field stars
    (blue and black data points) for each band.}
  \label{fig:Polmap}
\end{figure}

\begin{figure}[!htb]
  \centering
  \includegraphics[angle=0,width=8cm]{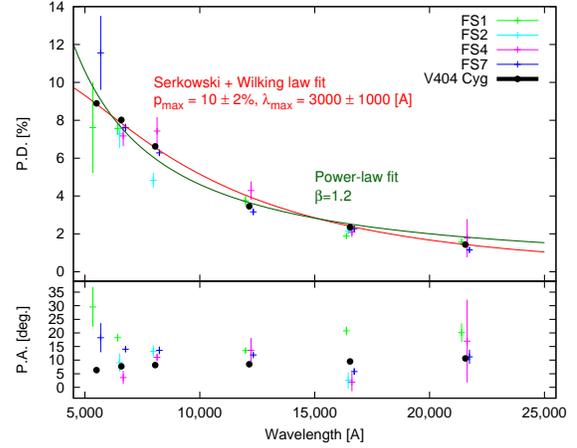}
  \caption{Multi-band polarization of V404~Cygni and field stars which showed high polarization degree.
    Top panel shows the dependence for polarization degree
    with wavelength and bottom panel shows the dependence for polarization angle with wavelength.
    Red line indicates the results of Serkowski-law + Wilking-law fitting, and green line indicates the
    results of power-law fitting (details are described in text).
    Note that in order to be easily seen, we shifted the wavelength value for field stars.}
  \label{fig:Polspec}
\end{figure}

Figure \ref{fig:Polmap} shows the multi-band polarization map around V404~Cygni. 
From this figure, the polarization angles (PAs) of field stars tends to
an average of 7 degree and a standard deviation of 15 degree
and V404~Cygni also shows a similar trend in the $R_{C}$-band. 
These trends of PA were also observed in the other band. 
Some field stars showed a small PD ($0\% < PD < 8\%$), but their PA tended to
align with the whole trend. 
These results indicate that V404~Cygni has interstellar polarization 
due to the interstellar medium between the Earth and sources. 
Figure \ref{fig:Polspec} shows the wavelength dependence of polarization
for V404~Cygni and field 
stars between the optical {\it V}-band and NIR $K_{s}$-band. 
For comparison, the wavelength dependence of polarization for field stars 
that have a high-polarization degree (FS1, FS2, FS4, and FS7) 
are also shown in Figure \ref{fig:Polspec}. 
PD and PA values for V404~Cygni are also listed in Table \ref{tab:Polspec}. 
The maximum PD of V404~Cygni is $PD=8.9 \pm 0.1$\% in the {\it V}-band, and 
PD decreased with wavelength, showing no clear peak down to 5500 \AA. 
This implies that the peak wavelength is located at shorter wavelengths ($< 5500$\AA). 
On the other hand, the PA was constant for all wavelengths, with an average value of 
PA $=8\pm 2$ degrees. 
V404~Cygni and all of the field stars with a high PD showed similar trends. 
This implies that the dust grains aligned by the interstellar magnetic field 
toward this region are generally smaller than the Galactic average 
\citep[e.g.,][]{2003dge..conf.....W,2014ApJ...795L...4K}. 
We adopted the Serkowski+Wilking law \citep{1975ApJ...196..261S} for the
wavelength dependence of PD for 
V404~Cygni, as described below, 
$PD_{\rm Ser} = p_{\rm max} \times
    \exp\left(-(1.66 \times10^{-4}\lambda_{\rm max}+0.01)\ln^2\frac{\lambda}{\lambda_{\rm max}}\right).$
Fitting shows $\lambda_{\rm max} = 3000 \pm 1000$ [${\rm \AA}$] and $p_{\max} = 10 \pm 2$\%. 
The $\chi^2_{\nu}$ value is $\chi^2_{\nu}/{\rm d.o.f} = 30.13/4$ for the fitting. 
The results of fitting implied a peak wavelength at $\lambda_{\rm max} = 3000$ [${\rm \AA}$], 
but the original data did not show a clear peak. Therefore, we used 
an upper limit of $\lambda_{\rm max} < 5500$ [\AA] as the peak wavelength for 
calculation of extinction. 
 
For comparison, we also adopted a simple power-law fitting for the PD dependence of wavelength, 
which are related to an analog of the ''IR polarization excess'' found in the Galactic ISP at 
longer wavelengths \citep[e.g.,][]{1990ApJ...348L..13N}, 
described as 
\begin{equation} 
PD_{\rm PL} = P_0 \lambda^{-\beta}, 
\end{equation} 
in Figure \ref{fig:Polspec}. 
The wavelength dependence of PD is well represented with $\beta = 1.2 \pm 0.1$ and 
$\chi^2_{\nu}/{\rm d.o,f} = 110.96/4$. 
We note that the value of $\beta=1.2$ obtained for the region of 
V404~Cygni is slightly shallow compared with the 
typical value of $\beta = 1.5-2.0$ for Galactic sources.

\begin{table}[!htb]
  \caption{PD and PA value for V404~Cygni}
  \centering
  \label{tab:Polspec}
  \begin{tabular}{lccc}\hline\hline
    Band        & Wavelength [\AA] & PD [\%] & PA [deg.]\\ \hline
    $V$         &  5504 & $8.9 \pm 0.1$ & $ 6.3 \pm 0.3$ \\ 
    $R_{\rm C}$ &  6587 & $8.0 \pm 0.1$ & $ 7.6 \pm 0.3$ \\
    $I_{\rm C}$ &  8059 & $6.6 \pm 0.1$ & $ 8.1 \pm 0.1$ \\
    $J$         & 12149 & $3.5 \pm 0.1$ & $ 8.4 \pm 0.1$ \\
    $H$         & 16539 & $2.3 \pm 0.1$ & $ 9.5 \pm 0.6$ \\
    $K_{s}$        & 21555 & $1.4 \pm 0.1$ & $10.6 \pm 0.7$ \\
    \hline
  \end{tabular}
\end{table}

\subsection{Estimation of Color excess}

\begin{figure}[!htb]
  \centering
  \includegraphics[angle=0,width=8cm]{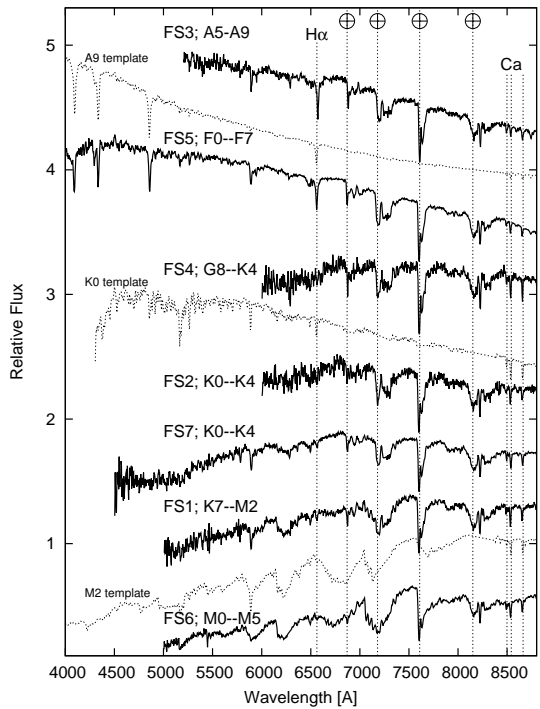}
  \caption{The optical spectra of field stars obtained by HOWPol.
    Correction of extinction are performed by estimated color excess.
    Dashed lines show the spectrum template for standard stars \citep{2000A&AS..143....9W}. 
    The mark of $\oplus$ indicate the teluric absorption lines.
    Summary of estimated spectral type and color excess are listed
    in Table \ref{tab:FSres}.}
  \label{fig:FSSpec}
\end{figure}

Figure \ref{fig:FSSpec} shows the optical spectra of field stars FS1-7 obtained by HOWPol. 
Comparing the patterns of dominant spectral features 
(e.g., H$_{\alpha}$, Ca IR triplet, TiO bands) with the 
template spectral atlas of standard stars \citep{1992ApJS...81..865S}, 
we estimated the spectral type of FS1-7. 
For comparison, some template spectra of standard stars are also shown in Figure \ref{fig:FSSpec}. 
Then, we measure the colour excess $E(R_{\rm C}-I_{\rm C}) = (R-I)_{observed} - (R-I)_{intrinsic}$,
using $(R_{\rm C}-I_{\rm C})_{\rm observed}$ obtained from our observations and
$(R_{\rm C}-I_{\rm C})_{\rm intrinsic}$ from well-studied field stars which have
the same spectral type as the stars in FS1-7 \citep{2000A&AS..143....9W}.
The same methods were used for the derivation of $E(B-V)$ with UVOT data. 
The uncertainties of color excess depended on the classification of spectral type. 
Finally, we adopted the correction of extinction for the optical spectra 
based on the estimated color excess. 
The measurements are summarized in Table \ref{tab:FSres}.

\begin{table*}[!htb]

  \caption{Summary of properties for field stars}
  \begin{center}
    \label{tab:FSres}
    \begin{tabular}{llccccc}\hline\hline
      Name & Spectral type$^1$ &Flux [mag.]$^2$ & P.D. [\%]$^3$& $R_{\rm C}-I_{\rm C}^4$ & $E(R_{\rm C}-I_{\rm C})^5$ &   $E(B-V)^6$\\ \hline
      FS1  & K7 -- M2 & 15.5 & $7.6 \pm 0.3$ & $1.77 \pm 0.02$ & $0.8 \pm 0.2$ & $ 1.0 \pm 0.3 $ \\
      FS2  & K0 -- K4 & 16.2 & $7.5 \pm 0.7$ & $1.49 \pm 0.02$ & $1.0 \pm 0.1$ & $ 1.6 \pm 0.5 $ \\ 
      FS3  & A5 -- A9 & 14.6 & $2.9 \pm 0.2$ & $0.67 \pm 0.02$ & $0.5 \pm 0.1$ & $ 0.8 \pm 0.1 $ \\
      FS4  & G8 -- K4 & 16.1 & $7.3 \pm 0.5$ & $1.75 \pm 0.02$ & $1.3 \pm 0.1$ & $ 0.8 \pm 0.6 $ \\
      FS5  & F0 -- F7 & 12.3 & $0.16\pm 0.05$& $0.31 \pm 0.02$ & $0.1 \pm 0.1$ & $ 0.2 \pm 0.1 $ \\
      FS6  & M0 -- M5 & 13.6 & $5.1 \pm 0.5$ & $2.47 \pm 0.02$ & $1.6 \pm 0.5$ & $ 1.3 \pm 0.1 $ \\
      FS7  & K0 -- K4 & 14.5 & $7.6 \pm 0.2$ & $1.60 \pm 0.02$ & $1.1 \pm 0.1$ & $ 1.5 \pm 0.1 $ \\
      \hline
    \end{tabular}
    \\
      {\footnotesize 1: Estimated spectral type, 
        2: Observed magnitude in the $R_{\rm C}$-band, 
        3: Observed optical polarization degree in the $R_{\rm C}$-band,
        4: Observed $R_{\rm C}-I_{\rm C}$ color,
        5: Estimated color excess $E(R_{\rm C}-I_{\rm C})$,
        6: Estimated color excess $E(B-V)$.
      }
  \end{center}
\end{table*}

\begin{figure*}[!htb]
  \centering
  \includegraphics[angle=0,width=15cm]{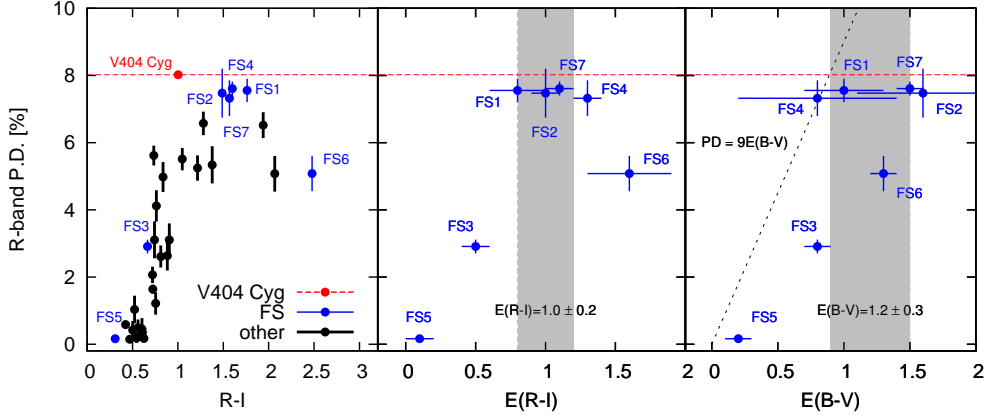}
  \caption{Left panel shows scatter plot for polarization degree and color $R-I$,
    central panel shows scatter plot for polarization degree and estimated color excess $E(R-I)$
    and right panel shows scatter plot for polarization degree and estimated color excess $E(B-V)$.
    Red data indicates the data of V404~Cygni, blue data indicate the data of Field stars (FS),
    and black data indicates the other field stars. In central and right panel, gray regions indicate
    the estimated $E(R-I)$ and $E(B-v)$ for V404~Cygni.
    Dashed line indicates the empirical maximum polarization degree as a function of $E(B-V)$
    \citep{2002ApJ...564..762F}.} 
  \label{fig:ext_PD}
\end{figure*}

Figure \ref{fig:ext_PD} shows a scatter plot for PD, color ($R_{\rm C}-I_{\rm C}$), 
estimated color excess $E(R_{\rm C}-I_{\rm C})$ and $E(B-V)$ for V404~Cygni and field stars. 
We also plotted all of the data points of field stars (not FS stars) in a scatter plot for 
color and polarization with black data points. 
From this figure, we can see that there is a general increase in PD with color and
color excess.

The field stars FS1, FS2, FS4, and FS7 show similar 
properties of PD and color excess (see Table \ref{tab:FSres}). 
From these results, we estimated the color excess of V404~Cygni. 
Using average value of color excess of the field stars FS1, FS2, FS4 and FS7,
we estimated the color excess of $E(R_{\rm C}-I_{\rm C}) = 1.0 \pm 0.2$ and
$E(B-V) = 1.2 \pm 0.3$ for the PD value of $PD = 8.0 \pm 0.1$\% for V404~Cygni. 
These values are consistent with the value derived by spectroscopic observations of 
A comparison star for V404~Cygni \citep{1993MNRAS.265..834C}.
We note that these color excess values are derived based on the PD value during the outburst.

\section{Discussion}

The aligned polarization angle and wavelength dependence of polarization implies 
that the polarization observed in V404~Cygni originated in the interstellar 
medium \citep[This is also discussed in][]{2016ApJ...823...35T}.
On the other hand, \cite{2016MNRAS.463.1822S} reported that
PD value in quiescent state is $PD = 7.41 \pm 0.32$\% on 2016 May 26.
This value is slightly low compared with the PD value in 2015 outburst and
several short time variability of PD which probably originated in the jet
during the 2015 outburst were also reported
\citep[e.g.,][]{2016MNRAS.463.1822S,2016arXiv160802764L}.
But variability of PD are relatively rare phenomenon during the outburst and
its variability amplitude of PD is small ($\Delta PD \sim 1-2\%$). 
Therefore, in this paper, we assumed that most of polarization originated in ISP.
In this section, we estimate the Galactic extinction and 
discuss the dust properties of V404~Cygni.

From optical spectroscopic observations, 
we identified the spectral type and determined the color excess 
of the field stars around V404~Cygni. 
The color excess and PD showed a good correlation, 
and we estimated the color excess of V404~Cygni as 
$E(R_{\rm C}-I_{\rm C}) = 1.0 \pm 0.2$ and $E(B-V) = 1.2 \pm 0.3$,
which corresponds to $A(V) = 3.7$ with $R_{\rm V} = 3.1$. 
We also measured the upper limit of peak wavelength for multi-band PD 
as $\lambda_{\rm max} < 5500$\AA. 
It is known that a relation between $\lambda_{\rm max}$ and $R_{\rm V}$ 
is described with the empirical formula 
$R_{\rm V} = 5.5 \times 10^{-4} \lambda_{\rm max}$ \citep{1975ApJ...196..261S}. 
We obtained the upper limit of $R_{\rm V} < 3.0$ and $A(V) < 3.6$ with 
$\lambda_{\rm max} < 5500$\AA. 
In addition, using another empirical relation of the maximum value of
$P(V)/A(V) < 0.03$ mag$^{-1}$ 
\citep[e.g.,][]{2008JQSRT.109.1527V}, where $P(V)$ is the PD value at $V$-band, 
we obtained the lower limit of $3.0 < A(V)$. 
Finally, we obtained an extinction value of $3.0 < A(V) < 3.6$. 
This extinction value is consistent with the value of $2.2 < A(V) < 3.3$
reported in \cite{2003MNRAS.346.1116S},
but slightly low compared with the value of $A(V)=4.0$ reported in
\cite{2009MNRAS.399.2239H}.
From X-ray observation, we also estimate the hydrogen column density via spectral fitting.
Using relation between hydrogen column density $N_H$ and $E(B-V)$ of
$N_H = (6.86\pm0.27) \times 10^{21} E(B-V)$ \citep{2009MNRAS.400.2050G},
$N_H = (8 \pm 2)\times 10^{21}$ cm$^{-2}$ is estimated with $E(B-V)=1.2$ and this value is
also consistent with X-ray observation of integrated hydrogen column density
of $N_H = 6-12 \times 10^{21}$ cm$^{-2}$ during the outburst in 2015 \citep{2016MNRAS.462.1834R}.
In general, measurements of the Galactic extinction for X-ray binary systems
are model-dependent \citep[e.g.,][]{2003MNRAS.346.1116S}. 
On the other hand, our method does not include the uncertainties caused by the emission model 
of a comparison of stars and disks as it independent from the accurate intrinsic luminosity 
of binary systems. 
Our method is applicable to other Galactic transients that have no intrinsic polarization 
in the optical band.

The typical value of the interstellar PA within 3 degrees 
of V404~Cygni is about $35 \pm 35$ degrees 
\citep[stellar polarization catalogs,][]{2000AJ....119..923H}. 
Compared with this perspective trend of polarization in the Galactic plane, 
it is implied that local ISP around V404~Cygni is not irrelevant to the global Galactic ISP. 
On the other hand, the maximum PD of the ISP in the Galactic plane 
within 3 degrees for V404~Cygni is about $PD = 3.69 \pm 0.18$ 
for HD~331976 at $\lambda \sim 5400$\AA. 
Of course, this may be due to a lack of measurement of stellar polarization in this region 
(we have only 22 sources in the stellar polarization catalog within 3 degrees of V404~Cygni). 
However, the measured local ISP of $PD = 8.9\pm0.1\%$ for V404~Cygni in {\it V}-band and
the color excess of  $E(B-V) = 1.2$ are among the highest values in the
Galaxy \citep{2002ApJ...564..762F}. 
The tendency of the bluer peak of polarization ($\lambda_{\rm max}<5500$\AA) 
suggested that the dust grains toward the V404~Cygni region are generally 
smaller than the Galactic average.

\begin{figure}[!htb]
  \centering
  \includegraphics[angle=0,width=8cm]{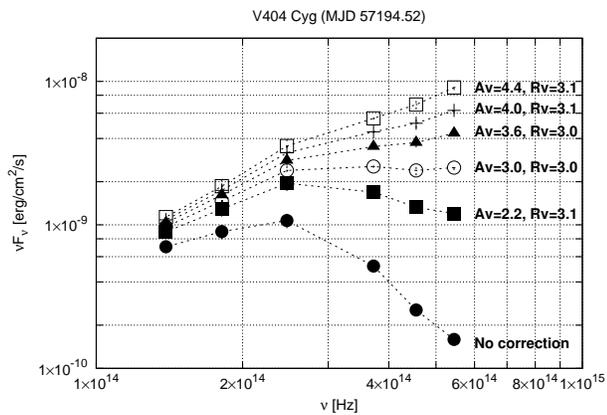}
  \caption{Quasi-simultaneous spectral energy distribution (SED)
    of V404~Cygni on MJD 57194.52
    with several correction of the Galactic extinction.
    Black filled circle data point indicates the SED without the Galactic extinction.
    Black box, open circle, black triangle, cross and open box data point corresponds
    to the SED with the Galactic extinction
    value of $A(V) = 2.2, R_{\rm V}=3.1, A(V) = 3.0, R_{\rm V}=3.0$, $A(V) = 3.6, R_{\rm V}=3.0,
    A(V) = 4.0, R_{\rm V}=3.1$ and $A(V) = 4.4, R_{\rm V}=3.1$ respectively.} 
  \label{fig:SED}
\end{figure}

Figure \ref{fig:SED} shows the quasi-simultaneous spectral energy distribution (SED) 
of V404~Cygni, with a correction for Galactic extinctions with several $A(V)$ and $R_{\rm V}$ values. 
The SED data were taken by HONIR within 30 minutes on MJD~57194. 
For comparison, we showed the SED with $A(V) = 2.2--4.4, R_{\rm V}=3.1$, which are often used for correction 
for V404~Cygni, and the SED with no correction in Figure \ref{fig:SED}. 
There is a break between the $J$ and $I_{\rm C}$ bands (corresponding to 
$2.5 \times 10^{14}$ Hz and $3.7 \times 10^{14}$ Hz, respectively) for 
the SED with $A(V) = 3.0, R_{\rm V}=3.0$. 
In the optical band ($>2.5 \times 10^{14}$ Hz), the SED with $A(V) = 3.0, R_{\rm V}=3.0$ 
shows a flat spectrum in the $\nu F_{\nu}$ regime. 
With the synchrotron emission model in the NIR and optical bands, this break indicates 
the break frequency due to synchrotron self absorption (SAA, defined as $\nu_{\rm SAA}$) 
which is an important value for estimating the magnetic field strength in the jet. 
The estimated $\nu_{\rm SAA}$ value of $2.5 \times 10^{14} < \nu_{\rm SAA} < 3.7 \times 10^{14}$ Hz 
is consistent with the $\nu_{\rm SAA}$ value assumed in \cite{2016ApJ...823...35T}.

\section*{Acknowledgement}
This work is supported by JSPS KAKENHI Grant Numbers 24000004.
This work is also supported by JSPS and NSF under the JSPS-NSF
Partnerships for International Research and Education (PIRE).
This work is also supported by the Optical and Nearinfrared
Astronomy Inter-University Cooperation Program
by the Ministry of Education, Culture,
Sports, Science and Technology of Japan.
M.W. was supported by a Grant-in-Aid for Young Scientists
(A) (25707007) from JSPS.

\end{document}